\documentclass[preprint,aps,amsmath]{revtex4}

\usepackage{amssymb}
\usepackage{graphicx}
\usepackage{subfigure}
\usepackage{upgreek}

\newcommand{\dd}{\mathrm{d}}
\newcommand{\bnabla}{\mbox{\boldmath$\nabla$}}
\newcommand{\corrsp}{\mathrel{\widehat{=}}}

\begin{document}

\title{Ergodicity of gyrofluid edge localised ideal ballooning modes}

\author{J. Peer and A. Kendl}
\address{Institut f\"ur Ionenphysik und Angewandte Physik, Universit\"at Innsbruck, Austria}
\author{B. D. Scott}
\address{Max-Planck-Institut f\"ur Plasmaphysik, Garching, Germany \vspace{1cm}}

\begin{abstract}
\vspace{1cm}
The magnetic field structure associated with edge localised ideal ballooning
mode (ELM) bursts is analysed by nonlinear gyrofluid computation.  The linear
growth phase is characterised by the formation of small scale magnetic islands.
Ergodic magnetic field regions develop near the end of the linear phase when
the instability starts to perturb the equilibrium profiles.  The nonlinear
blow-out gives rise to an ergodisation of the entire edge region.  The
time-dependent level of ergodicity is determined in terms of the mean radial
displacement of a magnetic field line.  The ergodicity decreases again during
the nonlinear turbulent phase of the blow-out in dependence on the degrading
plasma beta in the collapsing plasma pedestal profile.  

\vspace{7cm}
{\sl \noindent This is a preprint version of an article published
  in:\\ Plasma Physics and Controlled Fusion {\bf 55}, 015002 (2013).}
\end{abstract}

\maketitle

%------------------------------------------------------------------------
\section{Introduction}

The steep gradients related to the edge transport barrier in tokamak H-mode
plasmas facilitate the growth of edge localised modes (ELMs) involving
repetitive eruption of particles and energy \cite{zohm96,connor98,becoulet03}.
The largest and most vehement of such events, classified as ``Type-I'' ELMs,
are commonly associated with the onset of ideal or peeling ballooning modes in
edge pedestals \cite{snyder02,wilson06,snyder07}. 

In future large tokamak devices like ITER, the heat flux associated with type I ELMs
is estimated to seriously damage the plasma facing components (PFCs) and
methods for the suppression or at least effective mitigation of the
disruptions are essential for an economic steady state operation
\cite{federici03,loarte03,federici06}. 
One of the most promising ELM mitigation methods is the external
application of resonant magnetic perturbations (RMPs) which has been observed
to increase the ELM frequency and to reduce the heat load on the PFCs
\cite{evans05,evans06,liang07,suttrop11}. 
Models for the physics underlying the ELM mitigation by RMPs have been
developed~\cite{snyder07,tokar07,tokar08}. 
However, the successful mitigation even by nonresonant magnetic perturbations
renews questions about the acting mechanisms~\cite{suttrop11}. 

Numerical computations are an important tool to analyse the physics and mode
structure of ELMs. 
Ballooning ELM scenarios have been investigated in nonlinear simulations based on
magnetohydrodynamic (MHD) \cite{huysmans07,huysmans09,sugiyama10,pamela11},
two-fluid \cite{xu10,dudson11}, and gyrofluid \cite{kendl10} models.  

The magnetic structure has an essential part for development and
transport of ELMs. In addition, numerical investigations of the interaction
between ELMs and externally applied RMPs will require a detailed knowledge of
the parallel mode structure and the resulting magnetic flutter associated with
the ELM evolution in the perturbation-free case. 

The present work focuses on nonlinear gyrofluid computation of the dynamical
magnetic field structure associated with ideal ballooning ELMs.  The formation
of magnetic islands and the development of ergodic magnetic field regions, both
observed in MHD simulations \cite{huysmans07,huysmans09}, is reassessed with a
gyrofluid code that in addition allows the consistent treatment of the
small-scale turbulent blow-out \cite{kendl10,scott05,scott06}.  

It is found that an ideal ballooning ELM involves a distinct ergodisation of
the entire edge region. The decrease of the ergodicity in the turbulent
aftermath mainly depends on the evolution of plasma beta in the collapsing edge
region.

The paper is organized as follows: In
secs.~\ref{sec:gem}-\ref{sec:advmagtransp}, an overview of the model equations,
geometry and code is given, and suitable expressions for the evaluation of
ergodicity in the numerical results are defined.  The simulation setup and the
model for the initial H-mode state are discussed in sec.~\ref{sec:setup}.  The
results for the ELM induced magnetic field structure and the associated
formation of ergodic magnetic field regions are presented in
secs.~\ref{sec:magnstruct}-\ref{sec:transp}. In sec.~\ref{sec:summary}, the
results are summarized and discussed.

%------------------------------------------------------------------------
\section{Gyrofluid electromagnetic model and geometry}
\label{sec:gem}

The simulations presented in this work are performed with the nonlinear
gyrofluid electromagnetic model and code GEMR \cite{kendl10}. In the following
we review model equations and geometry.

GEMR includes six moment equations each for electrons and ions (labelled with
$z\in\{e,i\}$), which are coupled by a polarisation equation and an induction
equation \cite{scott05}.   The dependent variables are density $n_z$, parallel
velocity $u_{z\parallel}$, parallel temperature $T_{z\parallel}$, perpendicular
temperature $T_{z\perp}$, parallel component of the parallel heat flux
$q_{z\parallel\parallel}$, perpendicular component of the parallel heat flux
$q_{z\parallel\perp}$, electric potential $\phi$, and parallel magnetic
potential $A_\parallel$.  The full set of model equations are treated in
refs.~\cite{kendl10,scott05}. 

Here we use normalised quantities \cite{scott05}: The perpendicular spatial
scales are given in units of the minor plasma radius $a$. The time scale is
normalised by $a/c_{s0}$, where $c_{s0}=\sqrt{T_{e0}/M_i}$ is a reference
plasma sound speed. Here, $M_i$ denotes the ion mass and $T_{e0}$ is a
reference electron temperature.  The dependent variables are normalised by $n_z
\leftarrow n_z/n_{z0}$, $T_z \leftarrow T_z/T_{z0}$, $u_{z\parallel} \leftarrow
u_{z\parallel}/c_{s0}$, $q_{z\parallel} \leftarrow
q_{z\parallel}/(n_{z0}T_{z0}c_{s0})$, $\phi \leftarrow (e\phi)/T_{e0}$,
$A_\parallel \leftarrow A_\parallel/(\rho_{s0}\beta_{e0}B_0)$, where $n_{z0}$
represents a reference density, $T_{z0}$ is a reference temperature, $e$
denotes the elementary charge, $B_{0}$ represents the equilibrium magnetic flux
density, $\rho_{s0}=c\sqrt{M_iT_{e0}}/(eB_0)$ is the drift scale, and
$\beta_{e0}=4\pi p_{e0}/B_0^2$ is a reference value for the electron dynamical
beta. Here, $p_{e0}=n_{e0}T_{e0}$ denotes the reference electron pressure.  The
magnetic flux density is normalised by $B_0$. 

The model dynamically evolves the full profiles of the dependent variables,
where the inner (source) and outer (sink) radial boundaries are given by Neumann
and Dirichlet conditions, respectively. 
The computational domain includes an edge pedestal closed-flux-surface region with consistent
quasi-periodic parallel-perpendicular boundary conditions, and a
scrape-off-layer (SOL) where the parallel boundary conditions represent a Debye
sheath limiter placed at the bottom side of a circular torus
\cite{ribeiro05,ribeiro08}.

The main model parameters are the electron dynamical beta $\beta_{e0}$, the
normalised drift scale $\delta_0=\rho_{s0}/a$, and the collisionality
$\nu_{e0}=a/c_{s0}\tau_{e0}$, where $\tau_{e0}$ denotes a reference value for
the Braginskii electron collision time \cite{kendl10,scott05}.

The evolution of the profiles is self-consistently coupled to the magnetic
Shafranov equilibrium for circular flux surfaces.  Both the safety factor $q$
and the Shafranov shift are evolved in each time step \cite{scott06}.

The geometry is described in terms of field-aligned, unit-Jacobian Hamada
coordinates $(x,y_k,s)$ through
\begin{eqnarray}
   x &=& V = 2\pi^2R_0r^2, \\
   y_k &=& y-\alpha_k = q\theta-\zeta-\alpha_k, \\
   s &=& \theta \label{eq:ham}
\end{eqnarray}
where $V$ is the volume enclosed by the flux surface with major radius $R_0$
and minor radius $r$, and $\theta$ ($0\leq\theta<1$) and $\zeta$
($0\leq\zeta<1$) are the unit-cycle poloidal and toroidal Hamada angles (see
ref.~\cite{kendl10} for their definition). $V$ is given in units of $a^3$, and
$R_{0}$ and $r$ are normalised by $a$. In oder to avoid magnetic shear
deformation of grid cells, the $y$-coordinate is shifted by
$\alpha_k=q\theta_k+\Delta\alpha_k$, i.e. $\Delta\alpha_k$ is chosen to make
$\bnabla x$ and $\bnabla y_k$ locally orthogonal at $\theta=\theta_k$
\cite{scott01}.

The initial magnetic equilibrium is computed from a prescribed safety factor
profile $q_0(x)$.  The temporal evolution of the Shafranov shift and $q(x)$ are
determined by the Pfirsch-Schl\"uter current and the associated zonal ($m=n=0$)
component of $A_\parallel$.  The change of $q(x)$ in each time step is given by
\cite{scott06}
\begin{equation}
    \Delta\frac{1}{q} = 
	- {\delta_{0}\beta_{e0}R_{0} \over r} {\partial \over \partial r}
	\langle A_{\parallel}\rangle_{y,s}
    \label{eq:dq}
\end{equation}
where the brackets $\langle\ldots\rangle_{y,s}$ denote the zonal average over
$y$ and $s$, and the factors $\delta_0$ and $\beta_{e0}$ enter due to the
applied normalisation scheme.  The Shafranov shift is incorporated into the
coordinate grid by modifying the metric elements according to the s-$\alpha$
model.  The resulting relevant part of the coordinate metric is given by
\begin{eqnarray}
   g^{xx} &=& \bnabla x \cdot \bnabla x 
          = (V')^2 + \mathcal{O}(\varepsilon) 
	  = (2\pi)^4\left(R_0r\right)^2 + \mathcal{O}(\varepsilon) 
   \label{eq:gxx} \\
   g^{yy}_k &=& \bnabla y_k \cdot \bnabla y_k 
            = \frac{q^2}{(2\pi r)^2} + \mathcal{O}(\varepsilon) 
   \label{eq:gyy} \\
   g^{xy}_k &=& \bnabla x \cdot \bnabla y_k 
            = q'(\theta-\theta_k) - d'_s\sin(2\pi s) - \Delta\alpha_k' 
            \equiv 0 \quad \mathrm{at} \quad \theta=\theta_k 
   \label{eq:gxy}
\end{eqnarray}
where the prime denotes a derivative with respect to $r$, $g^{xx}$ and
$g^{xy}_k$ are given to lowest order in $\varepsilon=r/R_0$, and $d_s'$
represents the local magnetic shear given by the Pfirsch-Schl\"uter current.
In order to make $g^{xy}_k$ locally vanish at $\theta=\theta_k$, the shift in
the $y$-coordinate is defined as $\alpha_k = q\theta_k - d_s\sin\theta_k$,
where $d_s$ is given by
%---------- J.P.
%\begin{equation}
%    d_s = -{\delta_0\beta_{e0}q^2R_0 \over \pi r} {\partial \over \partial r}
%	\int dy \; A_\parallel\cos(2\pi s)
%    \label{eq:ds}
%\end{equation}
\begin{equation}
    d_s = -{\delta_0\beta_{e0}q^2R_0 \over \pi r} {\partial \over \partial r}
	\langle A_\parallel\cos(2\pi s)\rangle_{y,s}
    \label{eq:ds}
\end{equation}
%----------
The transformation ensures that magnetic field changes arising from the
axisymmetric component of $A_\parallel$ are placed on the coordinate grid.  To
avoid that this field component is considered twice, the axisymmetric part of
$A_\parallel$ is subtracted when the magnetic flutter is determined
\cite{scott06}.

%------------------------------------------------------------------------
\section{Magnetic field structure and ergodicity}

The focus of the present work is on the influence of edge localised ideal
ballooning mode bursts on the magnetic field structure.  The fluctuating
magnetic potential enters into the macroscopic magnetic equilibrium through its
zonal average via eq.~\ref{eq:dq} and sideband in eq.~\ref{eq:ds}.  The
magnetic shear is accordingly determined by $\hat s = q' (r/q)$ and the
Shafranov shift (eq.~\ref{eq:ds}).

The perpendicular magnetic flutter, which enters into the parallel
nonlinearities, results from the spatial variation of the non-axisymmetric part
$\widetilde A_{\parallel} = A_{\parallel} - \langle A_{\parallel}\rangle_{y}$ of
the parallel magnetic potential.  The magnetic fluctuations in direction of the
perpendicular unit vectors $\hat{\bf e}^x$ and $\hat{\bf e}^y_{k}$ are given by
\begin{eqnarray}
    \hat{B}^x &=& \widetilde{\bf B} \cdot \hat{\bf e}^x 
        = \left( \bnabla \times \widetilde{\bf A} \right) \cdot {\bnabla x \over \sqrt{g^{xx}}}
	= {\beta_{e0}\delta_{0} \over B^s\sqrt{g^{xx}}}
	{\partial \widetilde A_{\parallel} \over \partial y_{k}} 
    \label{eq:bx} \\
    \hat{B}^y_{k} &=& \widetilde{\bf B} \cdot \hat{\bf e}^y_{k} 
	= \left( \bnabla \times \widetilde{\bf A} \right) \cdot 
          {\bnabla y_{k} \over \sqrt{g^{yy}_{k}}}
	= - {\beta_{e0}\delta_{0} \over B^{s}\sqrt{g^{yy}_{k}}}
	 {\partial \widetilde A_{\parallel} \over \partial x}
    \label{eq:by}
\end{eqnarray}
where the factor $\beta_{e0}\delta_{0}$ results from the normalisation
scheme.  
%---------- J.P.
Eqs.~\ref{eq:bx} and \ref{eq:by} were derived by assuming $\widetilde{\bf
A}=A_{\parallel} {\bf b}$ and using the approximation
$\bnabla\times(A_{\parallel}{\bf b})\approx-{\bf b}\times\bnabla
A_{\parallel}$, where ${\bf b}={\bf B}/B$. The magnetic flutter field is
divergence free in good approximation. For the present work the corrections
resulting from the addition of lower order terms ensuring an exactly
divergence free magnetic flutter field were found to be negligible.
%----------

The contravariant components of the magnetic flutter allow to evaluate
the field line equation
\begin{equation}
    \frac{B^{x}}{\dd x} = \frac{B^{y}_{k}}{\dd y_{k}} = \frac{B^{s}}{\dd s}
    \label{eq:fieldline}
\end{equation}
which will be used to visualize the magnetic field structure in terms of
Poincar\'e plots.  

%---------- J.P.
Ergodic magnetic field regions develop when magnetic fluctuations destroy
the nested magnetic equilibrium surfaces.  The level of ergodicity can be
measured by the average radial field line displacement defined by \cite{finken99}
\begin{equation}
    \sigma(r_{0},l) = \frac{1}{N}\sum\limits_{i=1}^{N}\lvert r_{i}(l)-r_{0}\rvert
    \label{eq:raddispl}
\end{equation}
where $N$ denotes the number of considered field lines starting from a
reference flux surfaces with radius $r=r_{0}$, $l$ represents the length of a
field line measured with respect to its starting point on the reference flux
surface, and $r_{i}(l)$ gives the radial position of the $i$th field line after
a length $l$.
%----------

%------------------------------------------------------------------------
\section{Advective and magnetic transport}
\label{sec:advmagtransp}

Radial electromagnetic turbulent transport of heat and particles can occur by
fluid-like perpendicular $E\times B$ advection, or through parallel motion
along perpendicularly perturbed magnetic field lines.  The differential
formulation of the advective transport of ion density is given in field aligned
coordinates by
\begin{equation}
    \dd F_i^E
    = \left(n_i {\bf u}_E +
        T_{i\perp} {\bf w}_E \right) \cdot \dd {\bf S}
    = \left( n_i v_E^x + T_{i\perp} w_E^x \right )\dd y \dd s
    \label{eq:fie}
\end{equation}
where the area element $\dd {\bf S} = \dd y\dd y \bnabla x$ is oriented in radial direction.  
%---------- J.P., use normalized units everywhere
%${\bf v}_E = (c/B) {\bf b} \times \nabla \phi$ is the $E \times B$
%velocity, and ${\bf u}_E = (c/B) {\bf b} \times \nabla \phi_G$ and ${\bf w}_E
%= (c/B) {\bf b} \times \nabla \Omega_G$ include the 
%ion finite Larmor radius corrected potentials $\phi_G = \Gamma_1 \phi$ and $\Omega_G =
%\Gamma_2  \phi$. 
${\bf v}_E = \delta_{0}{\bf b} \times \bnabla \phi$ is the $E \times B$
velocity, and ${\bf u}_E = \delta_{0} {\bf b} \times \bnabla \phi_G$ and ${\bf w}_E
= \delta_{0} {\bf b} \times \bnabla \Omega_G$ include the 
ion finite Larmor radius (FLR) corrected potentials $\phi_G = \Gamma_1 \phi$
and $\Omega_G = \Gamma_2  \phi$. 
%----------
The screening operators $\Gamma_2 = b (\partial \Gamma_1 / \partial b)$ for
$b=k_{\perp}^2 \rho_i^2$ with $\Gamma_1 = \Gamma_0^{1/2}(b)$ are defined via
the gyroaveraging operator $\Gamma_0$ \cite{scott05}.  The gyro averaging and
screening operations are performed in Fourier space, and would in Pad\'e
approximation be given by $\Gamma_0^{1/2} \rightarrow (1+b/2)^{-1}$.  
The advective electron heat transport is 
\begin{equation}
    \dd Q_e^E
	= (0.5p_{e \parallel} + p_{e \perp}) {\bf v}_{E} \cdot \dd {\bf S}
        = (0.5p_{e \parallel} + p_{e \perp}) v_e^x\dd y \dd s
    \label{eq:qee}
\end{equation}
where $p_{e\parallel}=n_{e\parallel}+T_{e\parallel}$ and
$p_{e\perp}=n_{e\perp}+T_{e\perp}$ denote the linearised pressure in parallel
and perpendicular direction, respectively.

The magnetic flutter transport of ion density is given by
\begin{equation}
    \dd F_i^M
	= u_{i \parallel} {\bf b}\cdot\dd{\bf S} 
	= u_{i \parallel} b^x\dd y \dd s,
    \label{eq:fim}
\end{equation}
where $b^x$ denotes the radial component of the fluctuating magnetic field.
Correspondingly, the magnetic flutter transport of electron heat is defined by 
\begin{equation} 
    \dd Q_e^M
	= \left(q_{e \parallel \parallel} + q_{ e \parallel \perp} \right) {\bf b}\cdot\dd{\bf S}
	= \left(q_{e \parallel \parallel} + q_{ e \parallel \perp} \right) b^x\dd y\dd s.
    \label{eq:qem}
\end{equation}

The ion density transport in eq.~\ref{eq:fie} and \ref{eq:fim} is normalised by
$n_{i0} c_{s0}$. The electron heat transport in eq.~\ref{eq:qee} and
\ref{eq:qem} is given in units of $p_{e0}c_{s0}$.

%------------------------------------------------------------------------
\section{Model for initial H-mode state}
\label{sec:setup}

The onset of edge localised ideal ballooning modes is associated with the steep
pressure gradient ($\alpha_M=q^2R_{0}\lvert\bnabla\beta\rvert>\hat{s}$) in an
H-mode pedestal. The present edge turbulence model -- like any other available
first-principles based model -- is not able to obtain a self-sustained edge
transport barrier with experimentally realistic steep flow and pressure
profiles.  In contrast to MHD models, the presence of the finite-beta
ion-temperature-gradient instability at all values of beta and collisionality
remove the familiar MHD threshold from the model.  The H-mode edge state can
however be prescribed as an initial condition, and the blow-out of one ideal
ballooning event (destroying the transport barrier) can be computed from that.
The procedure to obtain and start from such an initial H-mode like state with
GEMR has been described in ref.~\cite{kendl10}.

The initial (reference) mid-pedestal values for density, temperature, and
magnetic field are motivated by an exemplary ASDEX Upgrade H-mode shot
(\#17151) \cite{horton05}, and are given as
$n_{e0}=n_{i0}=2.5\cdot10^{19}\,\mathrm{m}^{-3}$, $T_{e0}=300\,\mathrm{eV}$,
$T_{i0}=360\,\mathrm{eV}$, and $B_0=2\,\mathrm{T}$.  Major radius, minor
radius, and gradient lengths for density and temperature correspond to
$R_0=1.65\,\mathrm{m}$, $a=0.5\,\mathrm{m}$, $L_n=0.06\,\mathrm{m}$, and
$L_T=L_{\perp}=0.03\,\mathrm{m}$.  The radial simulation domain has an
extension of $\Delta r = 0.06\,\mathrm{m}$ around the separatrix located at
$r_0=1$,
%---------- J.P.
spanning both the pedestal and SOL regions.
%Note that due to the fact that the GEMR model is based on local
%midpedestal parameters (see sec.~\ref{sec:summary} for a discussion on this
%point), the use of a larger radial simulation domain would be increasingly
%inconsistent with the model assumptions.
%----------
The initial $q$-profile
is prescribed by $q_0=1.45+3.50\,(r/r_0)^2$, which corresponds to a reference
safety factor of $q_{0}(r_0)=4.95$ and a reference magnetic shear of
$\hat{s}_{0}(r_0)=1.41$.  The electron dynamical beta and the drift scale
resulting from the local reference parameters are $\beta_{e0} = 4 \cdot
10^{-4}$ and $\rho_{s0} = 1.25 \cdot 10^{-3} \,\mathrm{m}$.   The Braginskii
electron collision time is $\tau_{e0} = 2.56 \cdot 10^{-6} \,\mathrm{s}$.  

An evaluation of the MHD ballooning parameter gives
$\alpha_M=q^{2}R_{0}\lvert\bnabla\beta_{e}\rvert=2.3 >
\hat{s}_{0}$; thus the prescribed parameter set is expected to be ideal
ballooning unstable.
%---------- J.P.
The Lundquist number has the value $S=4\pi v_A L/(c^{2}\eta)=4.5\cdot10^{7}$,
where $v_A$ is the Alfv\'en speed, $\eta=0.51m_{e}\nu_{e0}/(n_{e0}e^{2})$
represents the plasma resistivity, and a characteristic length $L=R_{0}$ was
assumed. The high Lundquist number indicates the possibility to include low-$n$
ideal MHD modes \cite{dudson11}.
%----------

The initial pedestal profiles for density $n(r) = (n_0/2) (L_\perp/L_n)
g_0(r)$ and temperature $T(r) = (T_0/2) (L_\perp/L_{T}) g_0(r)$ are modelled by
\begin{equation} 
	g_0(r) = 1 - \sin \left( 2 \pi \; \frac{r-(r_0-\Delta r/4)}{\Delta r} \right)
	\quad\mathrm{for}\quad
	r_0-\Delta r/2 \leq r \leq r_0
\end{equation} 
For pre-processing of the initial state a reduced grid resolution is used and
the nonlinearities associated with the $E\times B$ advection and the parallel
derivative are shut off, while the profiles are gradually ramped up \cite{kendl10}.
A random turbulent pseudo-spectrum of density fluctuations with
a relative amplitude of $a_0=10^{-4}\rho_{s0}/2L_\perp$ was seeded on
the background only inside the closed flux surface region. Excluding seeding
on (and in the very vicinity of) the separatrix helps to reduce the onset spurious
growth of ion temperature gradient (ITG) separatrix modes during the further (full
scale) initial linear growth phase of the simulation. It was however found to
be necessary to additionally pin the initial profiles of ion density and (thus cutting off
neoclassical and parallel SOL transport around the separatrix) during the
first stage of the ideal ballooning mode growth phase to completely avoid
contamination by such unphysical separatrix ITG modes. 
%---------- J.P.
As soon as the ballooning instability enters the nonlinear regime and starts to
perturb the equilibrium quantities, all profiles are allowed to evolve
self-consistently.
%----------

%---------- J.P.
For this linear growth and nonlinear blow-out phase of the simulation a grid of
$64 \times 512 \times 16$ points in radial ($x$), perpendicular ($y$) and parallel
($s$) direction is used. The resolution in $(x,y)$ down to
$\rho_{s0}$ corresponding to $1.3\cdot10^{-3}\,\mathrm{m}$.  
The time step is set to $\Delta t=0.002\,a/c_{s}$, corresponding to 
$8.3\cdot10^{-9}\,\mathrm{s}$. 
Convergence during the MHD growth phase has been tested by doubling the
space and time resolution, with a deviation of less than 5\,\%.
%----------

%------------------------------------------------------------------------
\section{Computation of dynamical magnetic structure}
\label{sec:magnstruct}

The rapid nonlinear transition from ideal ballooning mode to turbulent
transport during a blow-out event has been characterised in
ref.~\cite{kendl10}.  In the following the perturbed magnetic field structure
of IBM bursts is analysed.  Both the changes in the magnetic equilibrium (i.e.
the variation of the $q$-profile) and the magnetic flutter (i.e. the magnetic
perturbations evolving from local current fluctuations) are considered.  

\begin{figure}
	\subfigure[$t=1$]{
	    \includegraphics[width=3.5cm]{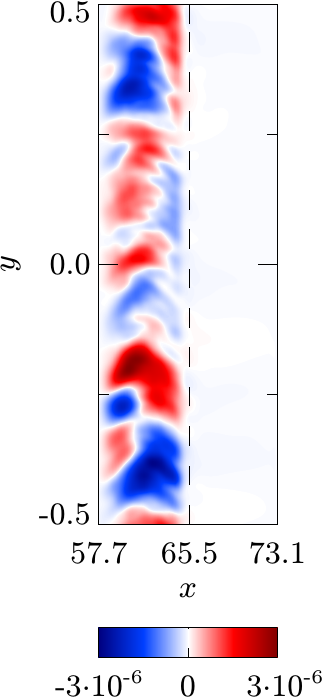}} \hfill
	\subfigure[$t=18$]{
	    \includegraphics[width=3.5cm]{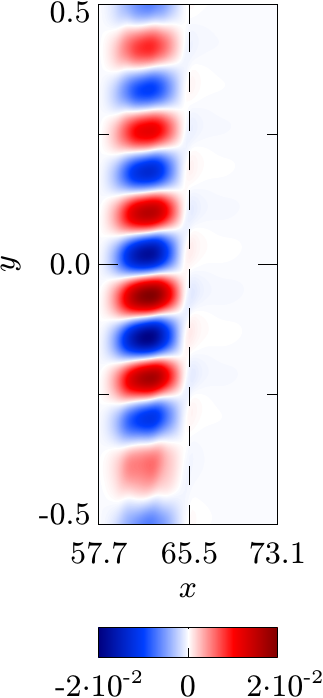}} \hfill
	\subfigure[$t=24$]{
	    \includegraphics[width=3.5cm]{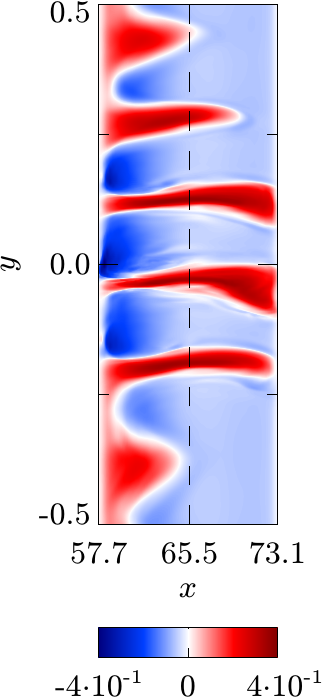}} \hfill
	\subfigure[$t=51$]{
	    \includegraphics[width=3.5cm]{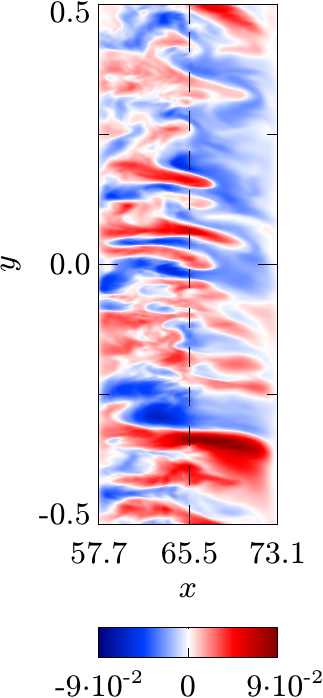}}
    \caption{(Colour online)
	Time evolution of an ideal ballooning mode burst illustrated by
	snapshots of electron density fluctuations in the outboard midplane
	($s=0$).  For visualisation of fluctuations  the toroidal mean has been
	subtracted  as $\widetilde{n}_e = n_e - \langle n_e \rangle_y$. The dashed
	lines mark the separatrix. Time and space scales are given in
	normalised units.
	}
    \label{fig:plctrsne}
\end{figure}

Fig.~\ref{fig:plctrsne} shows the time evolution of the density field
$n_{e}(x,y)$ of an IBM blow-out in the outboard midplane at various times of
the computation, starting from the initial conditions described above.  At
$t=1$ ($\corrsp4\,\upmu\mathrm{s}$) in fig.~\ref{fig:plctrsne} (a) the
fluctuations are randomly distributed within the closed flux surface region in
the left half space of the ($x,y$) domain. The ballooning instability starts to
grow where the pressure gradient is steepest, around $x \approx 61.3$.  The
linear growth phase, shown in (b) for $t=18$ ($\corrsp75\,\upmu\mathrm{s}$), is
characterised by periodic density fluctuations with mode number $n=6$.  It is
found that the dominant mode number depends on the pre-defined $q$-profile, but
for various computations with different initial conditions the mode numbers all
were in the range of $6 \leq n \leq8$.  After the peak growth phase nonlinear
saturation takes over. 
%---------- J.P.
Saturation occurs by energy transfer to drift-wave and ITG driven turbulence,
which has been demonstrated in the fluctuation spectra shown in fig.~6 of
ref.~\cite{kendl10}.
%----------
In (c) for $t=24$ ($\corrsp100\,\upmu\mathrm{s}$) the radial finger-like
interchange density perturbations show their most marked appearance.  In the
turbulent aftermath, shown in (d) at $t=51$ ($\corrsp210\,\upmu\mathrm{s}$),
the system transitions into a fully developed turbulent state with a mixed
drift wave and ITG character.
%---------- J.P.
The MHD mode starts perturbing the profiles at
$t\approx21$ ($\corrsp88\,\upmu\mathrm{s}$). The blowout phase (i.e. the 
induced erosion of density and temperature profiles) ends at $t\approx40$
($\corrsp170\,\upmu\mathrm{s}$) and has thus a duration of about
$80\,\upmu\mathrm{s}$.
%----------

\begin{figure}
	\subfigure[$q-q_{0}$]{
	    \includegraphics[width=7.5cm]{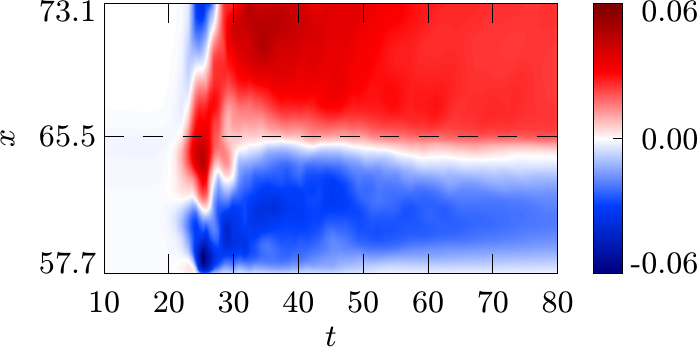}} \hfill
	\subfigure[$\hat{s}-\hat{s}_{0}$]{
	    \includegraphics[width=7.5cm]{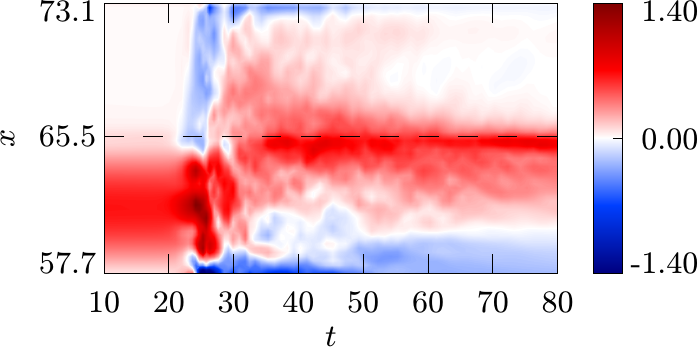}}
    \caption{(Colour online)
	Ideal ballooning ELM induced changes in the magnetic equilibrium.  The
	deviation of (a) the safety factor profile $q$ from the initially
	prescribed profile $q_{0}$ and (b) the corresponding deviation of the
	magnetic shear $\hat{s}$ from the initial shear $\hat{s}_{0}$ are
	shown. The dashed lines mark the separatrix. Time and space scales are
	given in normalised units.
	%Note that in fig.~(b) a small region adjacent to the inner radial boundary was
	%spared since boundary effects yielded unrealistic high shear values.
	}
    \label{fig:plequil}
\end{figure}

The effect of an IBM burst on the magnetic equilibrium ($n=0$) structure is
assessed via the change of the $q(x)$-profile according to eq.~\ref{eq:dq}.
Fig.~\ref{fig:plequil} shows the temporal evolution of changes in (a) safety
factor ($q(x,t)-q_0(x)$), and (b) magnetic shear ($\hat s(x,t)-\hat s_0(x)$).
Significant deviations from the initially prescribed values arise during the
blow-out phase and persist in the turbulent aftermath.

The difference between the time dependent $q$-profile and the initial profile
$q_0$ in (a) is in the range $-0.06 \leq (q-q_{0}) \leq 0.06$.
The nonlinear growth phase is characterised by positive deviations in the
central radial domain and negative deviations near both radial 
boundaries. In contrast, the turbulent aftermath exhibits positive deviations in
the SOL and around the separatrix, and negative deviations in the closed field
line region.
%---------- J.P.
The $q$-profile reflects the Shafranov shift: positive deviations from
$q_{0}$ indicate a decreasing, negative deviations an increasing shift.
%----------

The deviations of the magnetic shear in (b) are in the range
$-1.4 \leq (\hat{s}-\hat{s}_{0}) \leq 1.4$.  During the blow-out phase, the
shear deviations are essentially positive in the closed field line region
(except for the inner boundary) and essentially negative in the SOL.  In the
turbulent aftermath positive deviations dominate around the separatrix.
%---------- J.P.
The positive shear deviation until $t=21$ is due to the local magnetic shear
contribution (eq.~\ref{eq:ds}) in the initial equilibrium state.
%----------

\begin{figure}
	\subfigure[$t=1$]{
	    \includegraphics[width=7.5cm]{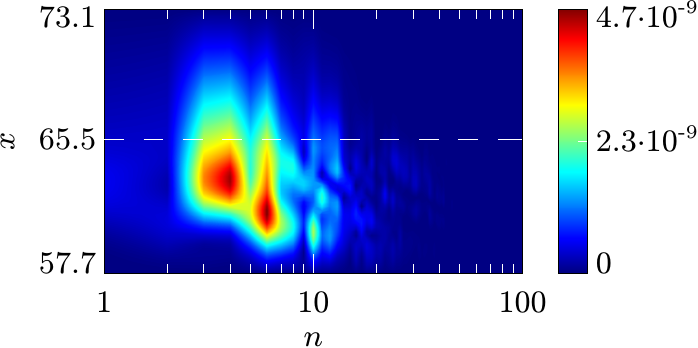}}\hfill
	\subfigure[$t=18$]{
	    \includegraphics[width=7.5cm]{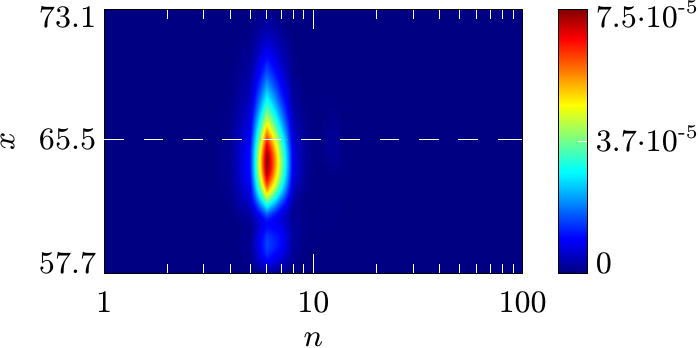}}
	\subfigure[$t=24$]{
	    \includegraphics[width=7.5cm]{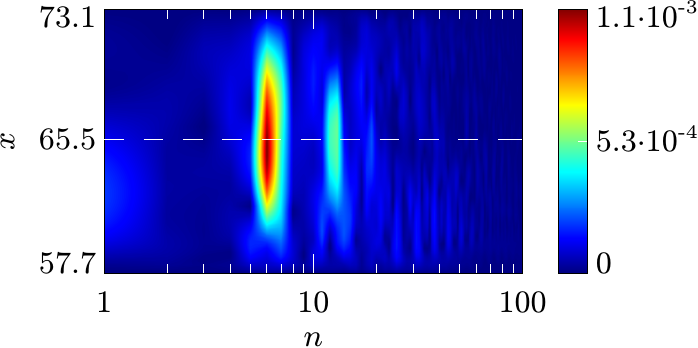}}\hfill
	\subfigure[$t=51$]{
	    \includegraphics[width=7.5cm]{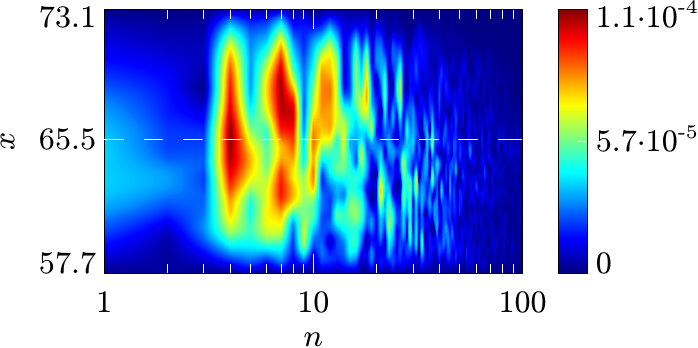}}
    \caption{(Colour online)
	Time evolution of magnetic flutter $\hat{B}^{x}$.  Toroidal mode number
	spectra in the outboard midplane ($s=0$) are shown at various times.
	The dashed lines mark the separatrix.  Time and space scales are given
	in normalised units.
	}
    \label{fig:plspecbx}
\end{figure}

\begin{figure}
	\subfigure[$t=1$]{
	    \includegraphics[width=7.5cm]{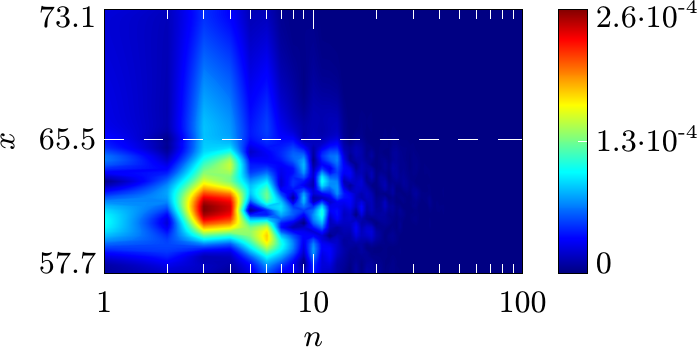}}\hfill
	\subfigure[$t=18$]{
	    \includegraphics[width=7.5cm]{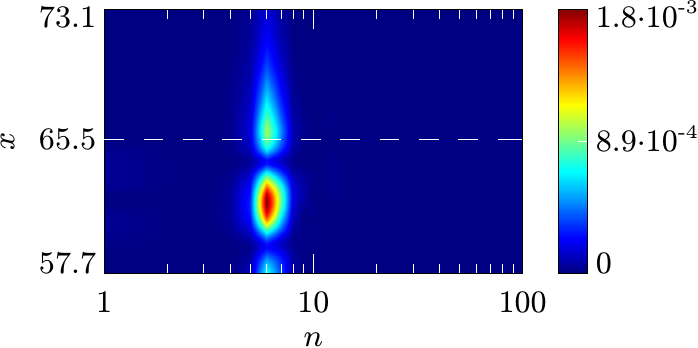}}
	\subfigure[$t=24$]{
	    \includegraphics[width=7.5cm]{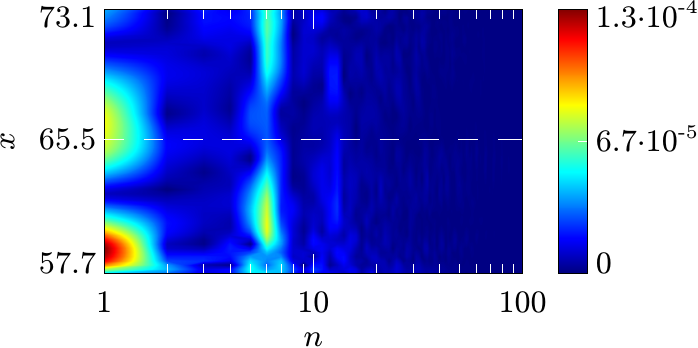}}\hfill
	\subfigure[$t=51$]{
	    \includegraphics[width=7.5cm]{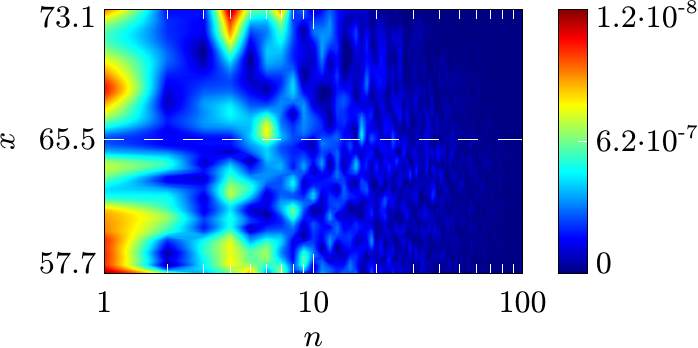}}
    \caption{ (Colour online)
	Time evolution of magnetic flutter $\hat{B}_{k}^{y}$.  Toroidal mode
	number spectra in the outboard midplane ($s=0$) are shown at various
	times.  The dashed lines mark the separatrix. Time and space scales are
	given in normalised units.
	}
    \label{fig:plspecby}
\end{figure}

%---------- J.P.
The equilibrium part of magnetic field changes has been determined by the
Pfirsch-Schl\"uter current and the associated axisymmetric $n=0$ toroidal mode
number component of the parallel magnetic potential. The blowout reduces
the Pfirsch-Schl\"uter equilibrium current by about 60\,\%.  Another
contribution to the dynamical magnetic structure is due to the perpendicular
magnetic perturbations resulting from the $n \geq 1$ components of $A_{\parallel}$. 
 The time variation of this magnetic flutter is examined by evaluating
 eqs.~\ref{eq:bx} and \ref{eq:by}. Figs.~\ref{fig:plspecbx} and 
\ref{fig:plspecby} show toroidal mode number spectra obtained from a Fourier
transform of the magnetic flutter on the outboard midplane ($s=0$).
%----------

The radial magnetic flutter $\hat{B}^{x}(x,t)$ in the $s=0$ plane is shown in
fig.~\ref{fig:plspecbx} at various times, comparable to the evolution of
structures in fig.~\ref{fig:plctrsne}.  Initially at (a) $t=1$ the
perturbations extend over a broad spectrum including mode numbers in the range
$2 \leq n \leq 20$ across the seeded confinement region.  The linear growth
phase at $t=18$ in (b) is dominated by perturbations with mode number $n=6$.
The neighbouring mode with $n=7$ is excited as well. The transition in (c) to
nonlinear saturation at $t=24$ involves a radial extension of the radial
magnetic flutter, reflecting the formation of interchange fingers far into the
SOL, and the second harmonic of the dominant mode becomes noticeably excited.
In (d) the turbulent aftermath at $t=51$ is characterised by nonlinear
excitation of multiple modes distributed over a broad range of mode numbers.
These signatures are also visible in the fluctuation spectrum as given in
fig.~6 of ref.~\cite{kendl10}. 

The evolution of the magnetic flutter component $\hat{B}^{y}_{k}(x,t)$ at
$s=0$, shown in fig.~\ref{fig:plspecby}, is similar to that of $\hat{B}^{x}$.
Remarkable differences concern the radial distribution of the perturbations
during the (b) linear growth phase and (c) nonlinear blow-out phase, and the
excitation of $n=1$ modes during the blow-out phase and the turbulent aftermath
in (d). 

The amplitude of the magnetic flutter increases until the IBM instability
saturates. The magnetic flutter is largest at $t=25$, where $|\hat{B}^{x}|
\approx |\hat{B}^{y}_{k}| \leq 10^{-2}$. The subsequent turbulent mixing
involves a decrease of the magnetic perturbation level.

\begin{figure}
	\subfigure[$\widetilde{J}_{\parallel}$]{
	    \includegraphics[width=3.5cm]{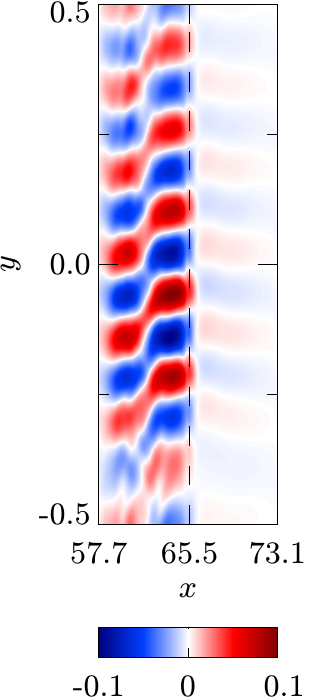}}\hfill
	\subfigure[$\widetilde{A}_{\parallel}$]{
	    \includegraphics[width=3.5cm]{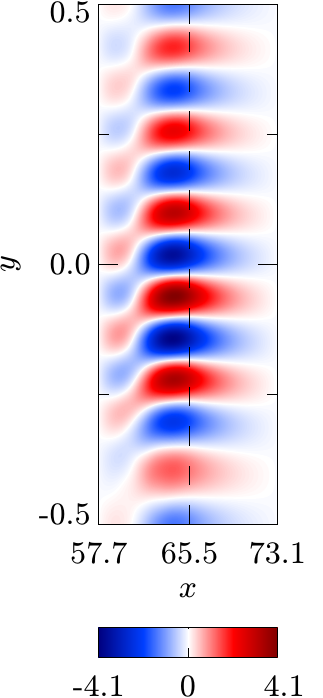}}\hfill
	\subfigure[$\hat{B}^{x}$]{
	    \includegraphics[width=3.5cm]{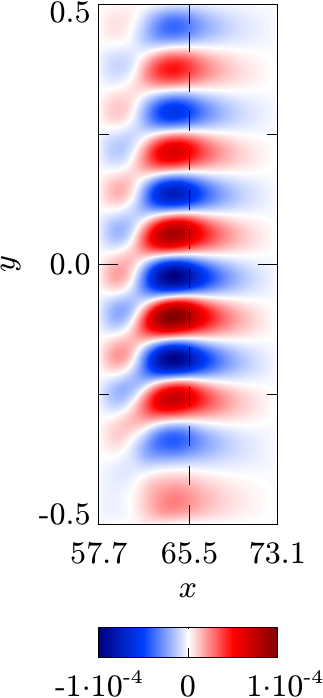}}\hfill
	\subfigure[$\hat{B}_{k}^{y}$]{
	    \includegraphics[width=3.5cm]{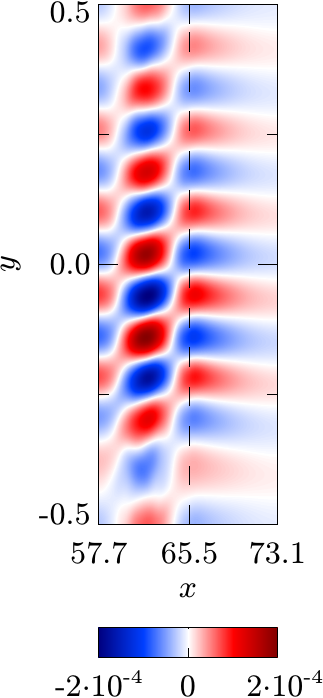}}
    \caption{ (Colour online)
	Fluctuations in (a) the parallel current and (b) the parallel
	magnetic potential, and the resulting (c, d) perpendicular magnetic
	flutter associated with the linear IBM growth phase at $t=18$.
	Fluctuations are shown by subtracting the toroidal mean
	($\widetilde{J}_{\parallel} = J_{\parallel} - \langle
	J_{\parallel}\rangle_{y}$, $\widetilde{A}_{\parallel} = A_{\parallel} -
	\langle A_{\parallel}\rangle_{y}$). The dashed lines mark the
	separatrix.  Time and space scales are given in normalised units.
	}
    \label{fig:plctrst18}
\end{figure}

The magnetic flutter in the linear growth phase is also well characterised by
the structure of the parallel current and the related parallel magnetic
potential.  Fig.~\ref{fig:plctrst18} shows the fluctuations in the parallel
current $J_{\parallel}(x,y)$, the related fluctuations in the parallel magnetic
potential $\widetilde A_{\parallel}(x,y)$, and the resulting magnetic flutter
terms $\hat B^x(x,y)$ and $\hat B^y_k(x,y)$ in the outboard midplane during the
linear growth phase at $t=18$.  The distinct periodic structure in
$J_{\parallel}$, shown in (a), reflects the growth of the $n=6$ density
perturbation.  The magnetic potential $\widetilde A_{\parallel}$ in (b) is
coupled to the current through Amp\`ere's law. The resulting structure in both
magnetic flutter components in (c) and (d) explains the radial distribution of
the magnetic flutter found in fig.~\ref{fig:plspecbx} (b) and
\ref{fig:plspecby} (b). 

%------------------------------------------------------------------------
\section{Magnetic islands and ergodicity}

\begin{figure}
	\subfigure[$t=18$]{
	    \includegraphics[width=15.5cm]{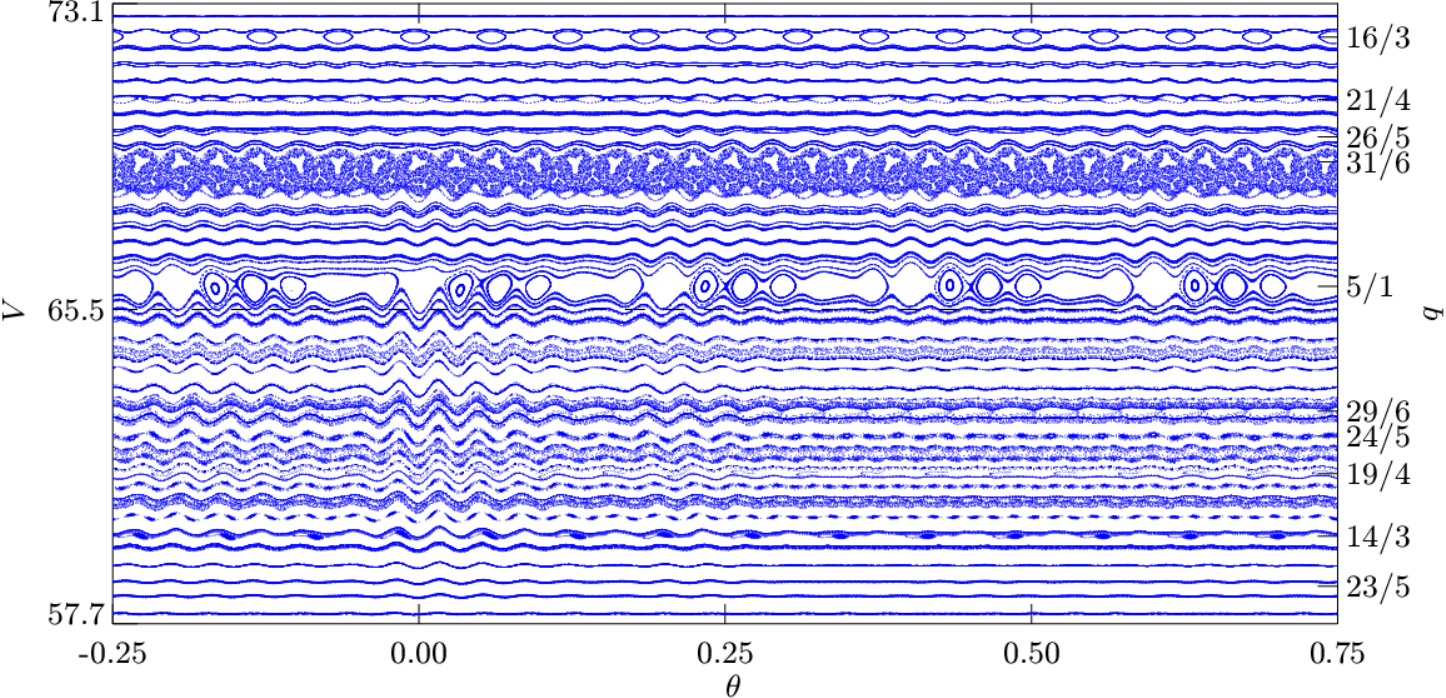}}
	\subfigure[$t=21$]{
	    \includegraphics[width=15.5cm]{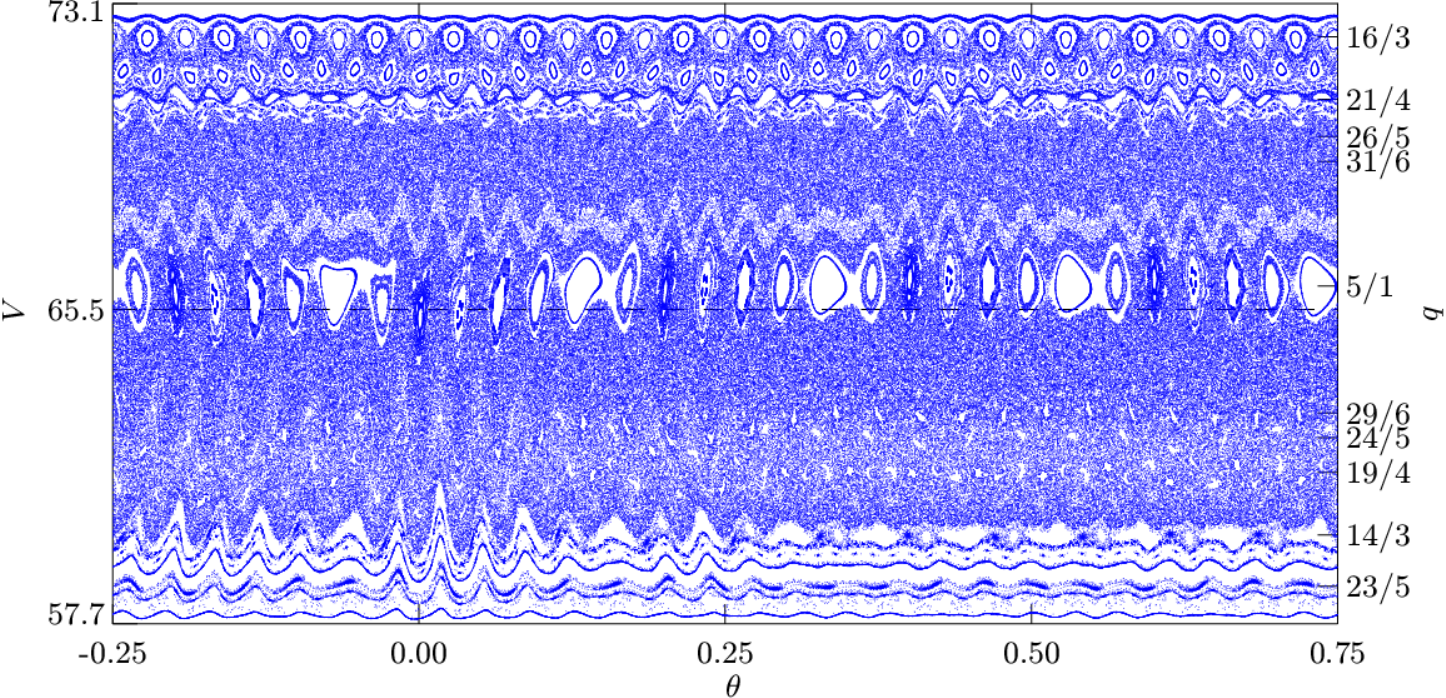}}
    \caption{
	Poincar\'e sections of the magnetic field lines illustrating the IBM
	induced transition from laminar magnetic flux surfaces at (a) $t=18$ to
	evolving ergodicity at (c) $t=21$. The dashed lines mark the
	separatrix. Time and space scales are given in normalised units.
	}
    \label{fig:plpcsec}
\end{figure}

The periodic structure in the magnetic flutter leads to the question whether the
linear growth of the IBM instability involves the formation of magnetic islands
at resonant rational flux surfaces. Integrating eq.~\ref{eq:fieldline}, 
the magnetic field structure can be visualised in terms of Poincar\'e sections. 
%---------- J.P.
Fig.~\ref{fig:plpcsec} shows Poincar\'e sections of the total simulation domain
at two simulation times.  For each plot, 160 field lines were traced over 4000
toroidal turns. Note that the requirement of a divergence-free magnetic field
means that, despite the implemented limiter, the magnetic field lines in the
SOL are closed. Hence, field lines can be traced over the entire radial
simulation domain.

In the linear phase at $t=18$, shown in fig.~\ref{fig:plpcsec} (a), the
magnetic flux surfaces in the closed-flux-surface region have an essentially
laminar structure. Several chains of small scale magnetic islands with island
widths $w\leq\rho_{s0}/2$ do not significantly perturb the flux surfaces. 
By contrast, applying the field line integration on the SOL (unphysically) exhibits larger,
partly overlapping islands of widths $w\leq 3 \rho_{s0}$. 
The largest islands in the SOL occur at rational surfaces
with $6\leq n\leq7$ (i.e. at $q=30/6,\,36/7,\,31/6,\,32/6$) and are thus
resonant with the linear IBM instability. The formation of magnetic islands in
the SOL during the linear IBM growth phase is actually not physically reasonable. 
The computed large amplitudes of islands in the SOL are likely an artefact
of the local $\delta f$ model, which overestimates $\beta$ and the
electromagnetic response during the linear growth phase in the SOL 
(see sec.~\ref{sec:summary} for a discussion on that point). 

At the transition to the nonlinear phase at $t=21$, shown in
fig.~\ref{fig:plpcsec} (b), most of the magnetic flux surfaces 
are destroyed and replaced by ergodic field regions. 
Note that the temporal onset of ergodicity coincides with the formation of
finger-like interchange perturbations.

\begin{figure}
	\subfigure[$t=18$]{
	    \includegraphics[width=15.5cm]{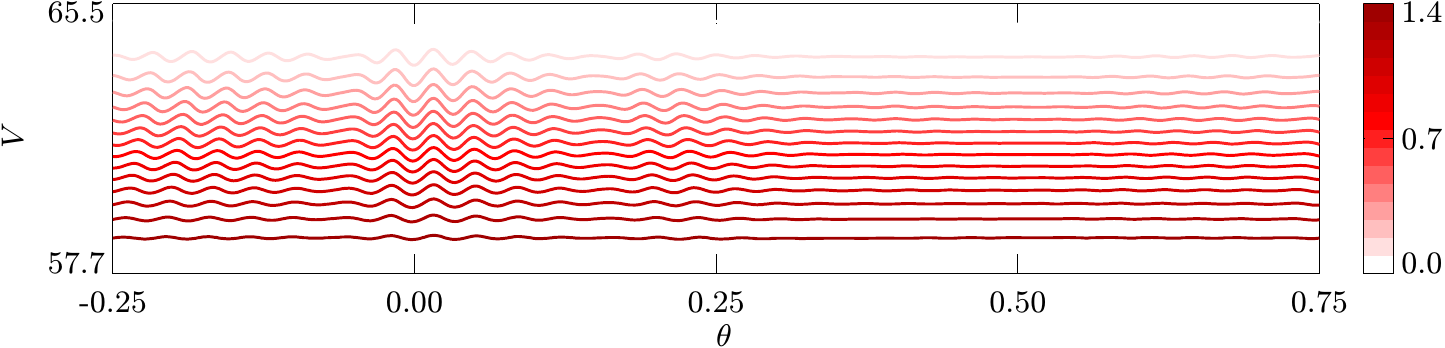}}
	\subfigure[$t=21$]{
	    \includegraphics[width=15.5cm]{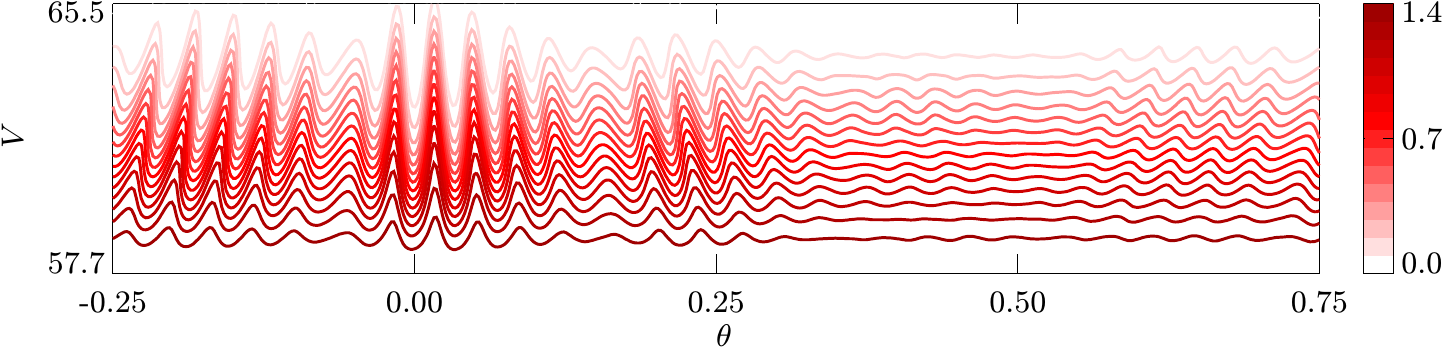}}
    \caption{
	Electron pressure contours in the poloidal plane. The closed-flux-surface region of
	figs.~\ref{fig:plpcsec} (a,b) is shown.  Time and space scales are
	given in normalised units.
	}
    \label{fig:plpolavpe}
\end{figure}

Fig.~\ref{fig:plpolavpe} shows electron pressure contours in the poloidal
plane.  In the linear phase at $t=18$, shown in fig.~\ref{fig:plpolavpe} (a),
the iso-pressure surfaces coincide with the magnetic flux surfaces found in
fig.~\ref{fig:plpcsec} (a). At $t=21$, the remaining laminar magnetic flux
surfaces still follow the electron pressure. Hence, as long as the pressure
fluctuations are small compared to the equilibrium profiles, the magnetic field
satisfies the frozen-in condition of ideal MHD. As soon as nonlinear saturation
takes over, resistive effects increase.

\begin{figure}
	\subfigure[]{
	    \includegraphics[width=7.5cm]{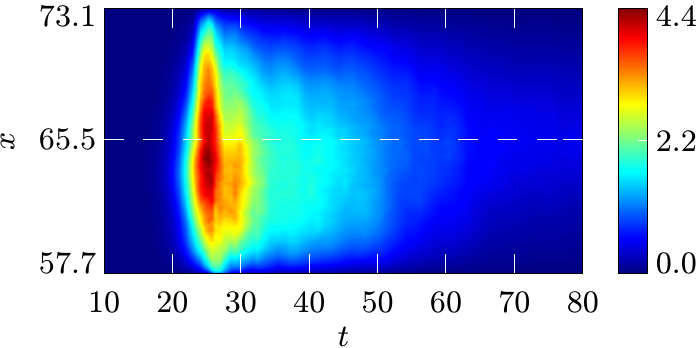}}\hfill
	\subfigure[]{
	    \includegraphics[width=7.5cm]{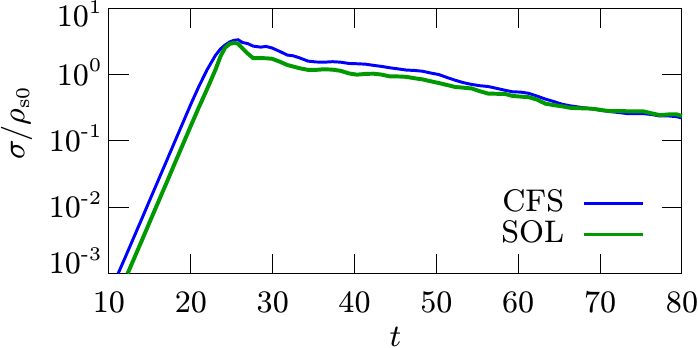}}
    \caption{ (Colour online)
	Mean radial field line displacement in units of the drift scale after
	one toroidal turn. The (a) radial variation on a linear colour scale
	and (b) the radially averaged displacement on a logarithmic scale is
	shown for the closed-flux-surface (CFS) and the scrape-off-layer (SOL)
	region.  The dashed line marks the separatrix. Time and space scales
	are given in normalised units.
	} 
    \label{fig:plddr}
\end{figure}

In order to quantify the time-dependent level of ergodicity, we evaluated the
average radial displacement of a magnetic field line, as defined by
eq.~\ref{eq:raddispl}.  Fig.~\ref{fig:plddr} shows the time dependent average
radial displacement of a magnetic field line after one toroidal turn in units
of the drift scale $\rho_{s0}$. The average was formed by tracing 1000 field
lines per flux surface.  Fig.~\ref{fig:plddr} (a) shows the displacement in
dependence on radial coordinate and time. In fig.~\ref{fig:plddr} (b) the
average displacement in both closed-flux-surface region and SOL is plotted on a logarithmic scale.
The linear growth phase is characterized by an exponential increase of the
field line displacement. The maximum displacement of
$\sigma\approx4\,\rho_{s0}$ coincides with the formation of the finger-like
interchange perturbations at $t\approx24$. Well after the ELM crash at $t=50$,
the displacement has reduced by 75\,\% with respect to its maximum value.

A comparison of the mean radial field line displacement with the Poincar\'e
sections shown in fig.~\ref{fig:plpcsec} shows that the IBM instability causes
ergodic field regions if $\sigma\gtrsim0.3\,\rho_{s0}$.  In the post-ELM
turbulent phase for $t>40$ a broad spectrum of modes contributes to the magnetic
flutter and the threshold for magnetic ergodicity is lower. Indeed, the ELM
blow-out induces an enduring ergodicity across the entire computation domain.
Even at $t=80$ ($\approx200\,\upmu\mathrm{s}$ after the ELM crash), where
$\sigma\approx0.2\,\rho_{s0}$, most of of the magnetic flux surfaces are
destroyed.

Here, the question arises of whether the ergodicity of the magnetic field is
even maintained in a saturated quasi L-mode-like post-ELM turbulent state.  As
the time scale of an ELM crash is much slower than the decrease of the profiles
associated with the turbulent transport, the computation of the IBM blow-out was
performed without density and heat sources.  In the following we discuss a
series of simulations including L-mode-like sources in the equations for
density and temperature.

The reduced pedestal density and temperature in an L-mode or post-ELM state
(compared to the H-mode transport barrier state) implies a reduction of the
electron dynamical beta and the drift scale, and an increase of the electron
collisionality.  Accordingly, we first investigated the scaling of the mean
radial field line displacement with the free model parameters $\beta_{e0}$,
$\nu_{e0}$, and $\delta_{0}$.  In a series of simulations, we varied
$\beta_{e0}$, $\nu_{e0}$, and $\delta_{0}$ parameter-inconsistently, and
evaluated $\sigma$ for the respective saturated states. As the sources control
the plasma fluctuation level, and consequently the amplitude of the magnetic
flutter, the dependence on the magnitude of the source flux was investigated as
well.

\begin{figure}
	\subfigure[Inconsistently varied parameters]{
	    \includegraphics[width=7.5cm]{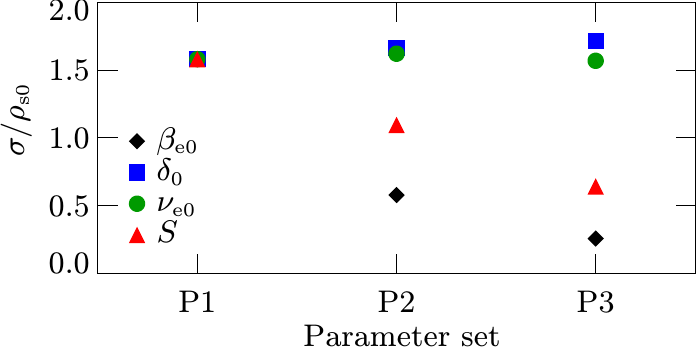}}\hfill
	\subfigure[Self-consistently varied parameters]{
	    \includegraphics[width=7.5cm]{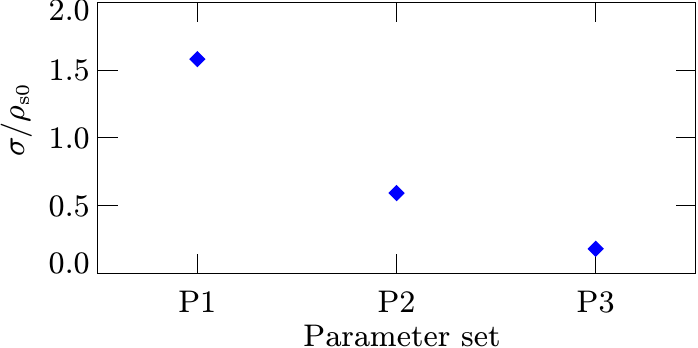}}
    \caption{
	Scaling of the mean radial field line displacement with the plasma beta
	$\beta_{e0}$, the drift parameter $\delta_{0}$, the electron
	collisionality $\nu_{e0}$, and the source level $S$.  Each point
	represents the average over a saturated turbulent L-mode-like state.
	The parameter set P1 corresponds to the nominal parameters. In P2 and
	P3, the parameters were computed from one half and one fourth of the
	nominal density and temperature profiles, respectively.  For (a) only
	one parameter was inconsistently varied, (b) shows parameter consistent
	simulations.  See text for details.
    }
    \label{fig:plddrvar}
\end{figure}

Fig.~\ref{fig:plddrvar} (a) illustrates the scaling of the mean radial field line
displacement with the electron dynamical beta $\beta_{e0}$, the electron
collisionality $\nu_{e0}$, the drift parameter $\delta_{0}$, and the source
level $S$.  Each point in the plot, represents an average over a time interval
$\Delta t = 400$ ($\corrsp1.7\,\mathrm{ms}$).  Except for the particular varied
parameters and the addition of sources $S^{nom}$, the nominal parameters
$\beta_{e0}^{nom}$, $\nu_{e0}^{nom}$, and $\delta_{0}^{nom}$ correspond to
those of the previous IBM case.  The sources comprise a positive flux of
density and heat acting near the inner radial boundary. The variation values of
the free model parameters were selected according to their density and
temperature dependence (i.e.  $\beta_{e0} \propto n_{e0}T_{e0}$, $\nu_{e0}
\propto n_{e0}T_{e0}^{-2}$, and $\delta_{0} \propto T_{e0}^{1/2}$). We
considered the cases with $n_{e0}=n_{e0}^{nom}$, $T_{e0}=T_{e0}^{nom}$ and
$S=S^{nom}$ (parameter set P1), $n_{e0}=n_{e0}^{nom}/2$,
$T_{e0}=T_{e0}^{nom}/2$ and $S=S^{nom}/2$ (parameter set P2), and
$n_{e0}=n_{e0}^{nom}/4$, $T_{e0}=T_{e0}^{nom}/4$ and $S=S^{nom}/4$ (parameter
set P3).  The variation of the plasma parameters shown in fig.~\ref{fig:plddrvar}
(a) shows that $\sigma$ decreases by 84\,\% if the electron dynamical beta is
decreased from $\beta_{e0}=\beta_{e0}^{nom}$ to
$\beta_{e0}=\beta_{{e}0}^{nom}/16$. On the other hand, a decrease of the drift
scale from $\delta_0=\delta_0^{nom}$ to $\delta_0=\delta_0^{nom}/2$ results in
a increase of $\sigma$ by 9\,\%.  The increase of the collisionality from
$\nu_{e0}=\nu_{e0}^{nom}$ to $\nu_{e0}=4\,\nu_{e0}^{nom}$ involves an
insignificant decrease of $\sigma$ by 1\,\%.  Finally, a reduction of the
source flux from $S=S^{nom}$ to $S=S^{nom}/4$ results in an decrease of
$\sigma$ by 59\,\%.

The above results indicate that the level of magnetic ergodicity is strongly
influenced by the values of the local plasma parameters, especially by the
plasma beta and the source level. A trend to lower ergodicity when going to
L-mode-like values is indicated. In order to verify this trend in
parameter-consistent simulations, we computed the mean radial field line
displacement for parameter-consistent, saturated L-mode-like states
characterized by the parameter sets P1, P2, and P3.  Fig.~\ref{fig:plddrvar} (b)
shows that $\sigma$ decreases by 88\,\% if the reference values for density and
temperature are parameter-consistently reduced to $n_{e0}=n_{e0}^{nom}/4$ and
$T_{e0}=T_{e0}^{nom}/4$ and the source level is lowered from $S=S^{nom}$ to
$S=S^{nom}/4$. Considering the Poincar\'e plots of the saturated state
characterized by the parameter set P3, we find a slightly ergodized magnetic
field. Thus, even L-mode-like turbulence can cause an ergodisation of the
magnetic field.  Note that this ergodicity is dynamically evolving with the
turbulence.
%----------

%------------------------------------------------------------------------
\section{Radial transport of ion density and electron heat}
\label{sec:transp}

As the IBM blow-out involves a distinct ergodicity of the magnetic field, an
increased magnetic transport of density and heat may be expected.  In order to
compare the magnetic transport (superscript $M$) with the $E\times B$ advective
transport (superscript $E$), the ion density transport $F_i$ and the electron
heat transport $Q_e$ are analysed. The motivation for considering $F_i$ instead
of $F_e$ is that the bulk of mass is transported by the ions.  On the other
hand, the higher mobility of the electrons implies that $Q_e^M \gg Q_i^M$, so
that only $Q_e$ is of interest.

\begin{figure}
	\subfigure[]{
	    \includegraphics[width=7.5cm]{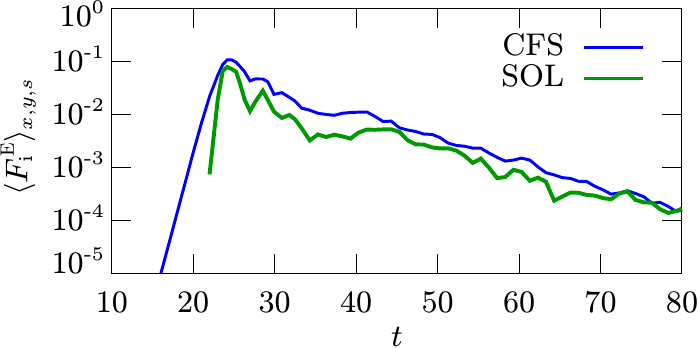}}\hfill
	\subfigure[]{
	    \includegraphics[width=7.5cm]{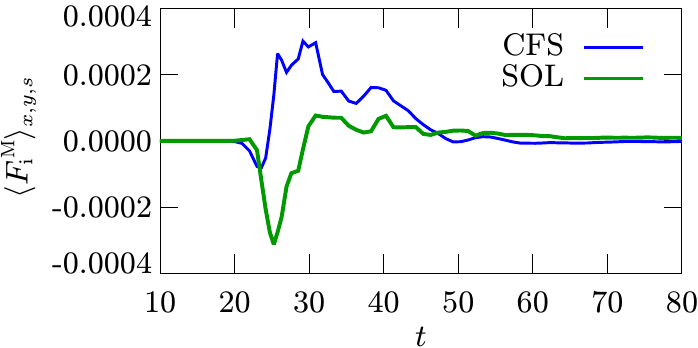}}\\
	\subfigure[]{
	    \includegraphics[width=7.5cm]{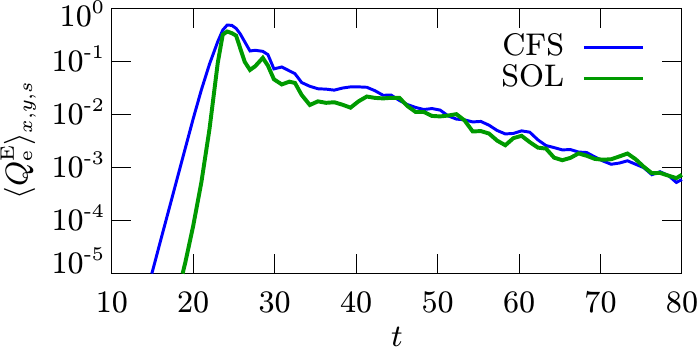}}\hfill
	\subfigure[]{
	    \includegraphics[width=7.5cm]{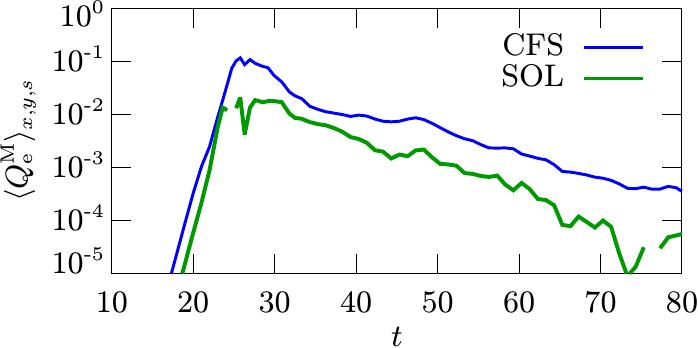}}
    \caption{
	Ideal ballooning ELM induced, volume averaged radial transport of (a,b)
	ion density and (c,d) electron heat.  The (a,c) $E\times B$ advective
	transport is compared to the (b,d) parallel magnetic transport.
	Closed-flux-surface (CFS) and scrape-off-layer (SOL) region are
	separately shown. Time and transport are given in normalised units.
	}
    \label{fig:pldgdw}
\end{figure}

Fig.~\ref{fig:pldgdw} shows the volume averaged transport quantities $F_i^M$,
$F_i^E$, $Q_e^M$ and $Q_e^E$ for both closed-flux-surface region and SOL. The ion density transport
shown in (a) and (b) is clearly dominated by $E\times B$ advection. The ratio
between magnetic transport and $E\times B$ transport is less than
$10^{-2}$, and the only relevant magnetic contributions are restricted to the
time interval $25 \lesssim t \lesssim 45$ around and shortly after the peak IBM mode
phase.

In contrast, the electron heat transport shown in (c) and (d) exhibits a
significant magnetic component. 
In the closed-flux-surface region after the peak IBM mode phase for
times $t \gtrsim 25$, the ratio between average magnetic and advective electron
heat transport is between the values $0.2\leq\langle
Q_{e}^{M}\rangle_{x,y,s}/\langle Q_{e}^{E}\rangle_{x,y,s}\leq 0.8$.  The
magnetic transport in the SOL is by approximately one order of magnitude
smaller than in the closed-flux-surface region so that the ratio $\langle
Q_{e}^{M}\rangle_{x,y,s}/\langle Q_{e}^{E}\rangle_{x,y,s}$ has values around
$0.1$ in the SOL.

%---------- J.P.
In the case of a saturated L-mode-state based on plasma parameters which
reflect post-ELM conditions (reduced plasma beta and drift parameter, increased
electron collisionality), the magnetic transport of electron heat is, on
average, by two orders of magnitude lower than the $E\times B$ advective
transport. Likewise, the magnetic density transport is by three orders of
magnitude smaller then the $E\times B$ transport. Thus, the only regime where
the magnetic transport substantially contributes to the total transport, is the
peak IBM mode phase and the following transition to a turbulent state at times
$24\lesssim t\lesssim40$.

\begin{figure}
	\includegraphics[width=7.5cm]{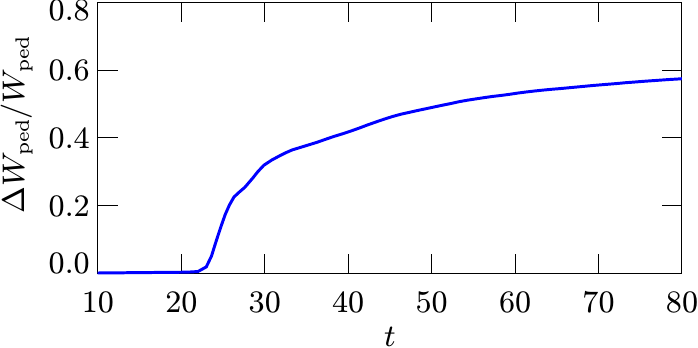}
    \caption{
	Ideal ballooning ELM induced loss of pedestal energy with respect to
	the equilibrium pedestal energy. The time is given in normalised units. 
	}
    \label{fig:plwped}
\end{figure}

The time-dependent total ELM energy loss, quantified by the loss fraction of
the equilibrium pedestal energy $W_{ped}=3/2(p_{e}+p_{i})$, is shown
in fig.~\ref{fig:plwped}. At $t=40$, which can be considered as the end of the
ELM crash, the energy loss amounts to about 42\,\%. At that time, the average
pedestal density of both electrons and ions has dropped by 41\,\% with respect
to the initial equilibrium state. The electron temperature at $t=40$ has
decreased by half and the ion temperature by a third. 
As the advective transport of electron and ion heat is nearly equal in magnitude, the larger
decrease of the electron temperature compared to the ion temperature can be
ascribed to the additional magnetic transport which is negligibly small for the
ions.
%----------

%------------------------------------------------------------------------
\section{Summary and discussion}
\label{sec:summary}

%---------- J.P.
The magnetic field structure associated with edge localised ideal ballooning
mode (IBM) bursts was investigated computationally using the nonlinear
gyrofluid electromagnetic model GEMR. The simulation setup was geared to
exemplary ASDEX Upgrade H-mode conditions. Both the IBM induced changes in the
magnetic equilibrium and the magnetic flutter due to local plasma fluctuations
were investigated.  The formation of magnetic islands and ergodic magnetic
field regions was visualized by Poincar\'e sections. In order to discuss the
level of magnetic ergodicity associated with the IBM blow-out and the
subsequent turbulent aftermath, the average radial displacement of a magnetic
field line was evaluated.  The level of ergodicity was investigated for
several saturated turbulent states in which the electron dynamical beta, the
electron collisionality, the drift scale and the magnitude of the source flux
were varied.  Furthermore, the mean radial field line displacement was
evaluated for parameter-consistent, saturated L-mode-like states.  Finally, the
volume averaged magnetic transport of ion density and electron heat was
compared to the corresponding $E\times B$ transport and the total ELM energy
loss was discussed.

The main results can be summarized as follows:
\begin{enumerate}
    \item The IBM induced changes in the safety factor profile amount up to
	1\,\% of the initially prescribed value $q_{0}$ ($0.99\lesssim
	q(x)/q_{0}(x) \lesssim1.01$). Due to the spatial variation of the
	safety factor profile (short-scale transitions from negative to
	positive deviations) and the additional local shear piece resulting
	from the Shafranov shift, the corresponding changes in the magnetic
	shear amount up to 100\,\% of the initially prescribed value
	$\hat{s}_{0}$ ($0\lesssim \hat{s}(x)/\hat{s}_{0}(x)\lesssim2$).  
    \item The time-dependent toroidal mode number spectra of both perpendicular
	magnetic flutter components reflect the evolution of the initially
	dominant $n=6$ mode. The different spatial variation of the magnetic
	flutter components is due to the structure in the pressure fluctuations
	and the related current fluctuations.  The magnitude of the magnetic
	flutter increases until the IBM instability nonlinearly saturates
	($|\hat{B}^{x}|\approx|\hat{B}^{y}_{k}|\lesssim10^{-2}$).
    \item The linear growth of the IBM instability involves the formation of
	small scale magnetic islands in the closed-flux-surface region
	($w\leq\rho_{s0}/2$). The laminar structure of the magnetic surfaces in
	the closed-flux-surface region is not significantly perturbed.  The
	formation of resonant magnetic islands in the SOL can be ascribed to
	the use of a local $\delta f$ model, which implies that the
	electromagnetic response in the SOL during the linear growth phase is
	highly overestimated.
    \item Ergodic magnetic field regions form near the end of the linear phase
	when the IBM instability starts to perturb the equilibrium profiles.
    \item The level of magnetic ergodicity (measured by the average radial field line
	displacement) increases until the IBM instability saturates.  The
	turbulent aftermath of the burst results in an enduring ergodicity
	across the simulation domain.
    \item Even in a saturated turbulent L-mode-like post-ELM state the magnetic
	field remains ergodized. However, the level of ergodicity decreases
	if the plasma parameters are adjusted to the reduced post-ELM profiles.
    \item The IBM induced magnetic transport of ion density can be neglected in
	comparison to the corresponding $E\times B$ transport. By contrast, the
	magnetic transport of electron heat can amount up to 80\,\% of the
	corresponding $E\times B$ transport. The magnetic transport of electron
	heat is significant during the blow-out phase and the subsequent
	transition to a turbulent state. In a saturated L-mode-like post ELM
	state the magnetic contribution to the electron heat transport is
	negligibly small.
    \item The total pedestal energy loss of the IBM burst amounts to about
	40\,\% of the equilibrium pedestal energy.
\end{enumerate}

Considering these points we can conclude that the ideal ballooning ELM scenario
involves an enduring ergodisation of the entire edge region. Moreover, the L-mode
like post-ELM state is also characterized by a non vanishing degree of magnetic
ergodicity. 

For correct interpretation of the results we have to discuss several
limitations of the present local (``$\delta f$'') turbulence model. The code
GEMR evolves the profiles as part of the dependent variables but the derivation
of the model equations is based on local parameters. The discrepancy between
radially varying profiles and inconsistent local plasma parameters increases
with the distance from the flux surface for which the plasma parameters were
defined.  This implies that especially in the SOL the deviation from the
consistent plasma parameters is large.  

For the simulations presented in this work this means that the plasma beta is
unrealistically high when considering the range around the separatrix and the SOL.
Hence, the electromagnetic response and, in particular, the magnetic flutter
are overestimated in the SOL.  Especially during the linear growth phase of the
IBM instability, where the profiles in the SOL are close to zero, the plasma
beta in the SOL is highly overestimated as self-consistent values would be
close to zero.  The formation of resonant magnetic islands and ergodic
field regions in the SOL during the linear growth phase can be ascribed to this
model inconsistency.

The discrepancy between profiles and plasma parameters also concerns the
transition from the peak IBM blow-out phase resulting from H-mode-like plasma
parameters to an L-mode-like turbulent post-ELM state which is characterized by
a decreased plasma beta. We have shown that the mean radial field line
displacement decreases up to one order of magnitude if the nominal simulation
parameters are changed to describe a saturated L-mode-like post-ELM state.

Our results that the magnetic heat transport during the ELM crash and the
subsequent turbulent state is always smaller than the heat transport by
$E\times B$ advection seems to disagree with the fact that the magnetic field
is ergodized at all times during and after the ELM blow-out. This apparent
contradiction can be solved by comparing the radial $E\times B$ velocity with
the radial velocity associated with the motion of an electron along a perturbed
magnetic field line.  Considering the mean radial field line displacement after
one toroidal turn and assuming an electron moving with thermal velocity, we
find that the average radial velocity associated with the magnetic flutter
during and after the peak blow-out phase is about one order of magnitude
smaller than the radial $E\times B$ velocity. Hence, the magnetic fraction of
the electron heat transport can be small although the magnetic field is
ergodised.  The point is that the magnetic flutter and the resulting ergodicity
is continuously changing with time such that an electron within one simulation
time step sees only a fraction of the magnetic snapshot visualized in a
Poincar\'e plot. Thus, an ergodic region in a Poincar\'e plot is not
necessarily indicative of a dominant magnetic transport.
%----------

The ELM induced ergodicity of the magnetic field was earlier investigated by
MHD simulations \cite{huysmans07,huysmans09,sugiyama10}. While MHD models more
readily allow for the implementation of a realistic X-point-geometry, the
present gyrofluid simulations (including turbulent scales) are based on a
simplified circular geometry. Still, the present results regarding the ELM
induced formation of an ergodic edge region as well as the associated magnetic
transport are in qualitative agreement with MHD results.  In the present
GEMR modelling the major novel results concern the small-scale dominated
turbulent aftermath immediately following the IBM burst.

\newpage

%------------------------------------------------------------------------
\section*{Acknowledgments}

We thank T.T. Ribeiro for valuable discussions on the model, code and results,
and F.P. Gennrich for routines for preparation of the initial state.  
This work was partly supported by the Austrian Science Fund (FWF) Y398; by the
Austrian Ministry of Science BMWF as part of the UniInfrastrukturprogramm of
the Forschungsplattform Scientific Computing at LFU Innsbruck; 
and by the European Commission under the Contract of Association between
EURATOM and \"OAW carried out within the framework of the European Fusion
Development Agreement (EFDA). The views and opinions expressed herein do not
necessarily reflect those of the European Commission.

%------------------------------------------------------------------------

%------------------------------------------------------------------------
% FIGURES

\newpage

\newpage

\newpage

\newpage

\newpage

\newpage

\newpage

\newpage

\newpage

\newpage

\newpage

\end{document}